\documentclass[preprint,showpacs,preprintnumbers,amsmath,amssymb]{revtex4}
\usepackage{epsfig}

\usepackage{graphicx}
\usepackage{dcolumn}
\usepackage{bm}
\def\lsim{\mathrel {\vcenter {\baselineskip 0pt \kern 0pt
    \hbox{$&lt;$} \kern 0pt \hbox{$\sim$} }}}
\def\gsim{\mathrel {\vcenter {\baselineskip 0pt \kern 0pt
    \hbox{$&gt;$} \kern 0pt \hbox{$\sim$} }}}

\newcommand{\U}{{\cal {U}}}

\begin{document}

\title{Unparticle Realization Through Continuous\\ Mass Scale Invariant Theories}

\author{$^1$N.G. Deshpande}
\email{desh@uoregon.edu}
\author{$^2$Xiao-Gang He}
\email{hexg@phys.ntu.edu.tw}
\affiliation{
$^1$Institute of Theoretical Sciences, University of Oregon, Eugene OR97401, USA
\\
$^2$Department of Physics
and Center for Theoretical Sciences, National Taiwan University,
Taipei, Taiwan, R.O.C.}

\date{\today}

\begin{abstract}
We consider scale invariant theories of continuous mass fields,
and show how interactions of these fields with the standard model
can reproduce unparticle interactions. There is no fixed point or
dimensional transmutation involved in this approach. We generalize
interactions of the standard model to multiple unparticles in this
formalism and explicitly work out some examples, in particular we
show that the product of two scalar unparticles behaves as a
normalized scalar unparticle with dimension equal to the sum of the two
composite unparticle dimensions. Extending the formalism to scale
invariant interactions of continuous mass fields, we calculate
three point function of unparticles.

\end{abstract}


\maketitle

Unparticle is an interesting idea proposed by Georgi\cite{Georgi:2007ek}, and is based on a scale
invariant sector weakly coupled to the Standard Model (SM). At lower energies the structure of the
scale invariant theory is assumed to have a fixed point in the coupling at a comparatively low
scale ($\sim$ TeV), below which by dimensional transmutation, operators emerge with non-integral
dimensions. As pointed out in Ref.\cite{Georgi:2007ek}, many interesting phenomena at TeV scale
emerge that can be understood purely from scaling properties of the unparticle operators. Although
this is completely satisfactory for phenomenology, much of the dynamics of the scale invariant sector
is mysterious, and the existence of a fixed point in the coupling parameter can only be hypothesized.

In this note we present a formulation that is based explicitly on a well defined Lagrangian which
possesses scale invariance. The Lagrangian involves continuous mass fields. One can now define
unparticle like local operators that couple to the SM. The unparticle properties emerge through
the choice of interactions. There is no fixed point or dimensional transmutation. The theory leads
to clear understanding of how unparticle exchange and phase space in the decay of SM particles arises.

One starting point is a free Lagrangian for a continuous mass scalar field
\begin{eqnarray}
L_0= {1\over 2} \int^\infty_0  [\partial_\mu \phi(x,s) \partial^\mu \phi(x,s) - s \phi^2(x,s)] ds\;.
\end{eqnarray}

The field equations are given by
\begin{eqnarray}
\partial_\mu {\partial L_0\over \partial \partial_\mu \phi(x,s) } = {\partial L_0 \over \partial \phi(x,s)}.
\end{eqnarray}

Using functional differentiation, we obtain
\begin{eqnarray}
(\partial_\mu \partial^\mu + s) \phi(x,s) = 0. \label{scaleeq}
\end{eqnarray}
These are infinite set of differential equations for all s from 0 to $\infty$.

On a historical note, we point out that such continuous mass
fields were studied long back by Thirring and others\cite{conti}
in the context of exactly soluble models. We also note that
continuous mass fields are also discussed by several
groups\cite{un-conti}  in context of unparticles, but in a somewhat
different spirit. Krasnikov in Ref.\cite{un-conti} has also considered continuous mass arising from a five-dimensional theory with broken Poincare invariance. We only consider ``s'' as a dimension 2 mass parameter. The theory can also be obtained as the continuum limit of infinite discrete mass fields.


We now discuss the scaling property of the theory under $x\to x' = \Lambda^{-1} x$. Since Lagrangian has dimension (mass)$^4$, the field $\phi(x,s)$ must have dimension zero.  To get its transformation property $\phi(x,s) \to \phi'(x',s)$ under scaling, we consider scaling property of the field equation (3). We have
\begin{eqnarray}
(\partial'_\mu \partial'^{\mu} + s) \phi' = (\Lambda^2 \partial_\mu \partial^\mu + s) \phi' = 0\;.
\end{eqnarray}
$\phi'$ is obviously a field of (mass)$^2 = s/\Lambda^2$. Thus taking into account that the field has dimension zero, we have under scaling,
\begin{eqnarray}
\phi(x,s) \to \phi'(x', s) = \phi(x, s/\Lambda^2)\;.
\end{eqnarray}

Since the mass $s/\Lambda^2$ is within the set of s from 0 to $\infty$, the transformed equations map on to the initial infinite set, and the theory is scale invariant. This was noted in Delgado et al in Ref.\cite{un-conti}.
To confirm scale invariance of the theory, we can explicitly see how the Lagrangian transforms under scaling. We have
\begin{eqnarray}
L_0 &\to & {1\over 2} \int^\infty_0 d s [\Lambda^2 \partial_\mu \phi(x,s/\Lambda^2)
\partial^{\mu} \phi(x, s/\Lambda^2) - s \phi^2(x, s/\Lambda^2)]\nonumber\\
&=& {1\over 2} {\Lambda^4} \int^{\infty}_0 [ \partial_\mu \phi(x, s/\Lambda^2) \partial^\mu \phi(x, s/\Lambda^2) -
{s\over \Lambda^2} \phi^2(x,s/\Lambda^2)]  d{s\over \Lambda^2} \;.
\end{eqnarray}
Changing integration variable to $s'=s/\Lambda^2$, we have $L_0 \to \Lambda^4 L_0$ and the action $S = \int d^4x L_0$ is invariant. Continuous mass thus restores the scale invariance that is broken by a theory with a discrete non-zero mass.

The field $\phi(x,s)$ in many ways is similar to a usual scalar
field, except that it is also labelled by a continuous mass
parameter $s$. We write a real $\phi(x,s)$ in its Fourier
representation as
\begin{eqnarray}
\phi(x,s)= \int {d^4p\over (2\pi)^4} 2\pi \delta(p^2 - s) \theta(p_0) [a(p,s) e^{-ipx} + a^\dagger(p,s) e^{ipx}].
\end{eqnarray}
Due to the fact that the $\phi(x,s)$ has a continuous mass parameter $s$, the quantization
rules for the creation and annihilation operators $a(k,s)$ and $a^\dagger(p,s)$ will be
different from that for a usual scalar filed. Appropriate generalization is the following
\begin{eqnarray}
[a(p,s), a^\dagger(k,s')] = (2\pi)^3 2 p_0 \delta^3(\vec p - \vec k)\delta(s-s').
\end{eqnarray}
Note that the dimension for $a$ and $a^\dagger$ is -2.

With the above quantization rules, we have
\begin{eqnarray}
<0|\phi(x,s)\phi(0,s')|0> = \int {d^4 p \over (2\pi)^4} e^{-ipx} 2\pi \delta(p^2-s)\delta(s-s').
\end{eqnarray}
and also the propagator is
\begin{eqnarray}
\int d^4 x e^{ipx} <0|T\phi(x,s)\phi(0,s')|0> = {i\over p^2 - s+i\epsilon} \delta(s-s').
\end{eqnarray}


A field with an arbitrary scaling dimension can now be constructed by convoluting the field $\phi(x,s)$
with a function $f(s)$ with a fixed
scaling dimension to have the following form
\begin{eqnarray}
\phi_\U(x) = \int^\infty_0 \phi(x,s) f(s) ds.
\end{eqnarray}
For  $f(s) = a_d (s)^{(d-2)/2}$, where
$a_d$ is an appropriately chosen normalization constant, $\phi_\U$ has scaling dimension d as can be seen by transforming $\phi(x,s) \to \phi(x, s/\Lambda^2)$ and changing integration variable to $s'=s /\Lambda^2$.

With the above definition, we have the following
\begin{eqnarray}
&&<0|\phi_\U(x) \phi_\U(0)|0> = \int {d^4p \over (2\pi)^4} e^{-ipx} 2\pi f^2(p^2)\;,\nonumber\\
&&\Pi = \int d^4x e^{ipx} <0|T\phi_\U(x) \phi_\U(0)|0> = \int {ds}
{i\over p^2-s+i\epsilon} f^2(s)\;.\label{free}
\end{eqnarray}
One can immediately identify the phase space $\rho(p^2)$ and propagator $\Pi$ for $\phi_\U$ to be
\begin{eqnarray}
&&\rho(p^2) = 2\pi f^2(p^2) = 2 \pi a^2_d (p^2)^{d-2}\;,\nonumber\\
&&\Pi = \int {ds} {i\over p^2-s+i\epsilon} f^2(s) ={(2\pi a_d^2
)\over 2 \sin(d\pi)}{i\over (-p^2)^{2-d}}\;.
\end{eqnarray}

Normalizing the constant $a_d$ as
\begin{eqnarray}
a^2_d = {A_d \over 2\pi}\;,\;\;\;\; A_d ={16\pi^{5/2}\over
(2\pi)^{2d}}{\Gamma(d+1/2)\over \Gamma(d-1)\Gamma(2d)}\;,
\end{eqnarray}
$\phi_\U$ has the same phase space and propagator as that defined
in Ref.\cite{Georgi:2007ek}, the unparticle operator. We also note
that fields obeying Eq. (\ref{free}) are called generalized free
fields\cite{green}. The special choice of $\rho$ makes them
transform with a unique scale dimension.

One can easily generalize the above formulation of unparticle to unparticles with different spins.
We display our results for
vector $A^\mu_\U$ and spinor $\psi_\U$ unparticles in the following.

For vector unparticle, we start  with
\begin{eqnarray}
&&L_0 =  \int^\infty_0  [-{1\over 4} F_{\mu\nu} F^{\mu\nu}
+{1\over 2} s A^\mu A_\mu] ds\;,
\end{eqnarray}
where $F_{\mu\nu} = \partial_\mu A_\nu - \partial_\nu A_\mu$. We note that presence of
$(mass)^2 = s$ means that the vector field is not gauge invariant.

$L_0$ is invariant under the scaling transformation:
$x \to \Lambda^{-1} x$, and $A^\mu \to A^\mu$. The vector unparticle
with dimension $d$ is defined by
\begin{eqnarray}
A_\U^\mu = \int^\infty_0  g(s) A^\mu(x,s) ds\;,\;\;\;\; g(s) = a_d
(s)^{(d-2)/2}\;,
\end{eqnarray}
and the phase space and propagator are given, in the transverse gauge, by
\begin{eqnarray}
&&\rho(p^2) = 2\pi g^2(p^2)(-g^{\mu\nu} + {p^\mu p^\nu\over p^2}),\nonumber\\
 &&\Pi = \int^\infty_0 {\rho(s)\over 2 \pi}{i\over p^2 - s +i\epsilon}ds
 = {A_d\over 2 \sin(d\pi)} {i\over (-p^2)^{2-d}}[-g^{\mu\nu} + {p^\mu p^\nu\over p^2}].
\end{eqnarray}

For spinor unparticle $\psi_\U$, we start with
\begin{eqnarray}
&&L_0 = \int^\infty_0  [\bar \psi i \gamma_\mu \partial^\mu \psi - \sqrt{s} \bar \psi \psi] ds\;,
\end{eqnarray}
which is invariant under the transformation:
$x \to \Lambda^{-1} x$, and $\psi \to \Lambda^{1/2} \psi$.
The spinor unparticle with dimension $d_\psi=d+1/2$, is given by
\begin{eqnarray}
\psi_\U = \int h(s) \psi(x,s) ds\;,\;\;\;\;h(s) = a_d
(s)^{(d-2)/2}\;,
\end{eqnarray}
and the phase space and propagator are given by
\begin{eqnarray}
&&\rho(p^2) = 2\pi h^2(p^2)(\gamma_\mu p^\mu + \sqrt{p^2}),\nonumber\\
 &&\Pi = \int^\infty_0 ds {\rho(s)\over 2 \pi}{i\over p^2 - s +i\epsilon}
= {A_d\over 2 \sin(d\pi)} {i\over (-p^2)^{2-d}} [\gamma_\mu p^\mu - i \mbox{ctg}(d\pi) \sqrt{p^2}] .
\end{eqnarray}

We note that if the vector field is a non-abelian massive field, it would violate scale invariance. This is because $F^a_{\mu\nu} = \partial_\mu A^a_\nu - \partial_\nu A^a_\mu + g f^{abc}A^b_\mu A^c_\nu$ has mixed transformation property under scaling since derivatives transform as $\Lambda$ and $f^{abc}$ as dimension zero. We can still have vector unparticles with non-trivial transformation under a group that do not have a continuous mass description.

We can consider operators that carry non trivial SM quantum numbers by replacing derivative
with covariant derivatives. This preserves the scale invariance of the theory for scalar and
spinor unparticles since covariant derivatives have the same dimension as the usual derivatives.
For spin one unparticle this is not possible because of additional self couplings.


We now comment on interactions of unparticles. Since one can now define unparticle like local operators,
the unparticle interaction with SM particles emerge through choice of interactions. The interaction of
unparticles with SM fields can be easily constructed from effective theory
point of view, using operators made of SM fields $O_{SM}$ and the unparticles $O_\U$ which can be one
of the $\phi_\U(x)$, $A^\mu_\U(x)$, or $\psi_\U(x)$ unparticle operators.

For one unparticle interaction with SM fields, the generic form is give by
\begin{eqnarray}
L_{eff} ={\lambda \over \Lambda_\U^{d_{SM} + d -4}} O_{SM} O_\U\;.
\end{eqnarray}
where $\Lambda_\U$ is a scale for the effective interaction and  $\lambda$ represents a dimensionless
coupling. $d$ and $d_{SM}$ are the dimensions of $O_\U$ and $O_{SM}$, respectively.

There are many ways unparticles can interact with SM sector. A set
of operators with SM operators have dimensions less or equal to 4
have been listed\cite{ Chen:2007qr} and many related phenomenology
have been discussed\cite{pheno}. We will not go into details about
related applications, except to point out that since we have
obtained the unparticle phase space and propagator, it is trivial
to carry out calculations for various applications, such as
unparticle production from colliders and decays, which go
completely parallel with those that have been considered in the
literature.  Instead we shall consider different processes.

Since now the unparticle operator $O_\U$ is treated as a local
operator, one can talk about multi-unparticles couplings among
themselves and also couplings to SM fields, such as interaction of
the form
\begin{eqnarray}
{\lambda_n\over \Lambda_\U^{d_1 + \cdot\cdot\cdot + d_n +d_{SM} - 4}}O_{SM}(O^1_\U\cdot \cdot \cdot O^n_\U)\;,
\end{eqnarray}
where $O_\U^i$ indicate an unparticle operator of dimension $d_i$.
When $d_{SM} = 0$, the above represents self-interactions of
unparticles.

Multi-unparticle interactions have some interesting properties. We
give a few examples in the following. Let us first consider the
propagator for the product of two scalar unparticles
$\phi_{\U_3}(x) = \phi_{\U_1}(x)\phi_{\U_2}(x)$ where
\begin{eqnarray}
&&\phi_{\U_i}(x) = \int ds f_i(s) \phi_i(x,s)\;,\;\;f_i(s) =
a_{d_i} s^{(d_i -2)/2}.
\end{eqnarray}
Note that the same $\phi_i(x,s)= \phi(x,s)$ can be used to
construct unparticles of different dimensions by convoluting a
different $f_i(s)$.

The propagator for $\phi_{\U_3}(x)$ is defined by $\Pi = \int d^4x
e^{ipx} <0|T\phi_{\U_3}(x) \phi_{\U_3}(0)|0>$. We have, using Wick
contraction,
\begin{eqnarray}
&&<0|T\phi_{\U_3}(x) \phi_{\U_3}(0)|0>
= <0|T\phi_{\U_1}(x) \phi_{\U_1}(0)|0><0|T\phi_{\U_2}(x) \phi_{\U_2}(0)|0>\nonumber\\
&&+ <0|T\phi_{\U_1}(x) \phi_{\U_2}(0)|0><0|T\phi_{\U_2}(x) \phi_{\U_1}(0)|0>\nonumber\\
&&=\int {d^4p_1\over (2\pi)^4} e^{-ip_1 x} \int d s_1
{if^2_1(s_1)\over p^2_1 -s_1 +i\epsilon} \int {d^4p_2\over
(2\pi)^4} e^{-ip_2 x} \int d s_2 {if^2_2(s_2)\over p^2_2 -s_2
+i\epsilon}.
\end{eqnarray}
Here we consider the case with $\phi_1\neq \phi_2$ so that the
cross term is zero. We will discuss the result for the same
$\phi_i=\phi(x,s)$ later.

Carrying out integrations for x and $p_i$ for $\Pi$, $\Pi$ can be
written as
\begin{eqnarray}
\Pi &=&\int^\infty_0 {\rho(s)\over 2 \pi} {i\over p^2 - s
+i\epsilon}\;,
\end{eqnarray}
with
\begin{eqnarray}
\rho(s) &=& \int^s_0 ds_1 \int^{(\sqrt{s}-\sqrt{s_1})^2}_0 ds_2
{1\over 8 \pi} f^2_1(s_1)f^2_2(s_2){1\over s} (s^2 - 2s (s_1+s_2) + (s_1-s_2)^2)^{1/2}\nonumber\\
&=&{a^2_{d_1}a^2_{d_2}\over 8 \pi} s^{d_1+d_2-2}\int^1_0 d x
\int^{(1-\sqrt{x})^2}_0 dy x^{d_1-2}y^{d_2-2}
(1-2(x+y)+(x-y)^2)^{1/2}\nonumber\\
&=&{a^2_{d_1}a^2_{d_2}\over 8 \pi} s^{d_1+d_2-2}
(d_1-1)(d_2-1)(d_1+d_2-1) {\Gamma^2(d_1-1)\Gamma^2(d_2-1)\over
\Gamma^2(d_1+d_2)}
\end{eqnarray}
Inserting $a^2_{d} = A_d/2\pi$ and using $\Gamma(2d) = \pi^{-1/2}
2^{2d-1}\Gamma(d+1/2)\Gamma(d)$, the above expression can be
written as
\begin{eqnarray}
\rho(s) = s^{d_1+d_2-2} {16\pi^{5/2}\over
(2\pi)^{2(d_1+d_2)}}{\Gamma(d_1+d_2+1/2)\over
\Gamma(d_1+d_2-1)\Gamma(2(d_1+d_2))}.
\end{eqnarray}
This is the phase space for a unparticle of dimension $d_3 =
d_1+d_2$.

We therefore have shown that $\phi_{\U_3}$ is an unparticle with
dimension $d_3=d_1+d_2$. The normalization $A_d$ is something
deeper than just convenience\cite{Georgi:2007ek}. Had another
normalization been used, the product of two scalar unparticle
would not be a new unparticle with dimension equal to the sum of
the two unparticles with the correct  normalization. The
self similarity of unparticle dictates the normalization.

For the case $\phi_{1} = \phi_{2}$, the cross term will also
contribute the same amount, but the total should be divided by
$2!$ to get the right normalization, in another words,
$\phi_{\U_3}$ should be written as
$\phi_{\U_1}\phi_{\U_2}/\sqrt{2!}$. One gets $\phi_{\U_3}$ to be
an unparticle of dimension $d_1+d_2$. The above discussions can be
easily generalized to any number of scalar unparticle product.
With proper permutation normalization, the product is an
unparticle with the dimension equal to the sum of the composite
unparticles. Products involving spinor and vector unparticles will
be more complicated,and we shall discuss them in detail in a
future publication.

As a further important application of Eq. (8), we calculate the three
unparticle vertex function defined by
\begin{eqnarray}
V(p^2_1, p^2_2, p^2_3) = \int d^4x e^{ip_1 x} d^4y
e^{ip_2y}<0|T(\phi_\U(x) \phi_\U(y) \phi(0))|0>,
\end{eqnarray}
where $p_3 = p_1 +p_2$.

From general scaling argument it follows that it has dimension
$3d-8$, and is invariant function of three variables $p^2_1$,
$p^2_2$ and $p^2_3$. Further, it is symmetric under exchanges
between $p_1$, $p_2$ and $p_3$. However scaling alone is not
sufficient to determine this function, as we shall see.

We first evaluate the time ordered product $T_3= <0|T(\phi_\U(x)
\phi_\U(y) \phi_\U(0)|0>$. We have
\begin{eqnarray}
T_3 = \int^\infty_0 ds_1ds_2ds_3
f(s_1)f(s_2)f(s_3)<0|T(\phi(x,s_1)\phi(y,s_2)\phi(0,s_3))|0>.
\end{eqnarray}

With Lagrangian in Eq.(1), since there are no interactions, the
integral is obviously zero, and there is no three point function.
We introduce scale invariant interactions of the continuous mass
fields so as to have non-vanishing $T$ product. It is sufficient
to introduce terms of $\phi^3$ type. The idea is to introduce some
dynamics that is also scale invariant, but at the same time, we
only consider tree level consequences of such a theory. Deeper
questions like renormalizability of such a theory are beyond the
scope of this paper.

One possible $\phi^3$ scale invariant interaction is
\begin{eqnarray}
L_{\lambda} = {\lambda\over 3!} \int {ds_1 ds_2 ds_3\over
(s_1s_2s_3)^{1/3} }\phi(x,s_1) \phi(x,s_2)\phi(x,s_3)\;.
\end{eqnarray}
Another possibility is
\begin{eqnarray}
L_{g} = {g\over 3!}\int^\infty_0 s ds \phi^3(x,s).
\end{eqnarray}

We note that the modifications to equations of motion from the above two interactions are
respectively
\begin{eqnarray}
\partial_\mu \partial^\mu \phi(x,s) + s \phi(x,s) = {\lambda\over 2 s^{1/3}}\int {ds_1 ds_2\over (s_1s_2)^{1/3}}\phi(x,s_1)\phi(x,s_2),
\end{eqnarray}
and
\begin{eqnarray}
(\partial_\mu \partial^\mu + s) \phi(x,s) = {g\over 2} s \phi^2(x,s)\;.
\end{eqnarray}

Both equations under scale transformation map within the infinite set of equations, as can be verified. The first is a integro-differential equation. Such equations have been considered previously \cite{conti}, where a model is solved exactly in the case of bilinear interactions. Difference between the above two forms can be understood if one goes to the discrete limit of the theory.

We evaluate the time ordered product $t_3 =
<0|T(\phi(x,s_1)\phi(y,s_2)\phi(0,s_3)|0>$ in the lowest order
perturbation theory and find using $L_\lambda$ interaction,
\begin{eqnarray}
t_3 &=& \lambda \int d^4z \int {ds'_1ds'_2 ds'_3\over (s'_1s'_2
s'_3)^{1/3}}<0|T\phi(x,s_1)\phi(z,s'_1)|0>\nonumber\\
&\times&
<0|T\phi(y,s_2)\phi(z,s'_2)|0><0|T\phi(0,s_3)\phi(z,s'_3)|0>\;.
\end{eqnarray}

Using
\begin{eqnarray}
<0|T(\phi(x,s)\phi(z,s')|0> = \int {d^4q\over (2\pi)^4}
e^{-iq(x-z)} {i \over q^2 -s+i\epsilon}\delta(s-s'),
\end{eqnarray}
we get
\begin{eqnarray}
t_3 = -i\lambda \int {d^4q_1\over (2\pi)^4} {d^2q_2\over (2\pi)^4}
{e^{-i(q_1 x +q_2 y)}\over (s_1s_2s_3)^{1/3}}{1\over (q^2_1-s_1
+i\epsilon) (q^2_2 - s_2 +i\epsilon)(q^2_3 - s_3 +i\epsilon)}\;,
\end{eqnarray}
where $q_3 = q_1+q_2$.

We now get for $T$ product of unparticles
\begin{eqnarray}
V(p^2_1,p^2_2,p^2_3) = -i \lambda \int
{f(s_1)f(s_2)f(s_3)ds_1ds_2ds_3 \over (s_1s_2s_3)^{1/3} (p_1^2
-s_1 + i\epsilon)(p^2_2 -s_2 +i\epsilon) (p_3^2 - s_3
+i\epsilon)}\;.
\end{eqnarray}
Using the formula,
\begin{eqnarray}
\int {f(s)ds \over s^{1/3} (p^2 - s +i\epsilon)} = \int {a_d
s^{(d-2)/2} ds \over s^{1/3}(p^2-s + i\epsilon)} = {a_d \pi \over
\sin((d/2-4/3)\pi)}{i \over (-p^2)^{-d/2+4/3}}\;.
\end{eqnarray}

Defining a new constant $\lambda' = \lambda
(a_d\pi/\sin((d/2-4/3)\pi))^3$, we have
\begin{eqnarray}
V(p^2_1,p^2_2,p^2_3) = -i \lambda' {1\over (-p^2_1)^{-d/2+4/3}
(-p^2_2)^{-d/2+4/3}(-p^2_3)^{-d/2+4/3}}\;.
\end{eqnarray}
This expression has the correct dimensions and fulfills all the
symmetry requirements.

A similar computation with $L_g$ interaction gives
\begin{eqnarray}
V(p^2_1,p^2_2,p^2_3) &=& -i g'[ {1\over (p^2_1 -
p^2_2)(p^2_1-p^2_3)} {1\over (-p^2_1)^{2-3d/2}} \nonumber\\
&+& {1\over (p^2_2 - p^2_1)(p^2_2-p^2_3)} {1\over
(-p^2_2)^{2-3d/2}} + {1\over (p^2_3 - p^2_1)(p^2_3-p^2_2)} {1\over
(-p^2_3)^{2-3d/2}}]\;,
\end{eqnarray}
where $g' = g a_d^3\pi/\sin(3d\pi/2)$. This expression also
fulfills all the requirements of dimensions and symmetry.

Note that in Ref.\cite{feng} Feng et al. calculate the three point
function assuming conformal invariance. Their starting point is
\begin{eqnarray}
<0|T(\phi_\U(x)\phi_\U(y)\phi_\U(0))|0> = C {1\over |x|^d}{1\over
|y|^d }{1\over |x-y|^d}\;.
\end{eqnarray}

However this form does not seem unique if only scale invariance is
imposed, for example one can multiply this by a dimensionless
function of $|x|/|y|$ and $|x-y|/|y|$. Another example for the
right hand side is $[1/|x|^{3d} + 1/|y|^{3d} + 1/|x-y|^{3d}]$. Our
expressions are simple in momentum space, but complicated in
coordinate space, while Ref.\cite{feng} has a complicated form in
momentum space.

The simplest form for the three point function in momentum space
is in Eq. (34). It is possible to probe the three point function
experimentally. Feng et al. have discussed various signals for
three point function. When standard model particles couple to
unparticles, it is possible to get signals that depend on the
explicit form of the vertex. For example if we have a coupling of
the type $\bar e e \phi_\U$, we can get events of the type $e^+e^-
\to e^+ e^- + e^+ e^-$ where the energy distribution of $e^+ e^-$
pairs will depend on the explicit three point function. Study of
such signals will be very useful, and we shall pursue it is future
publication.

We can extend our analysis to three point functions involving
scalar, spinor and vector unparticles. We have to add interactions
of continuous mass spinor and vector fields that preserve scale
invariance. As an example, we can add $L_{fh} = \int ds \sqrt{s}
[f \bar \psi(x,s) \psi(x,s) \phi(x,s) + h \bar \psi(x,s)
\gamma_\mu \psi(x,s) A^\mu(x,s)]$ or $L_{f'h'} = \int
(ds_1ds_2ds_3/ \sqrt{s_1s_2s_3}) [f' \bar \psi(x,s_1) \psi(x,s_2)
\phi(x,s_3) + h' \bar \psi(x,s_1) \gamma_\mu \psi(x,s_2)
A^\mu(x,s_3)]$. Consequences of such interactions will be pursued
in future publications.

One can also easily generalize to unparticles with SM gauge interactions by assuming that
the $\phi(x,s)$, and $\psi(x,s)$ to have non-trivial SM quantum numbers. The end results are
that the unparticle operators $\phi_\U$, and $\psi_\U$ have the same SM quantum numbers as $\phi(x,s)$,
and $\phi(x,s)$, respectively. When taking derivatives, one should take the covariant derivative as
would have to be done for usual particles. As pointed out earlier that to preserve scale invariance,
the vector operator cannot be non-abelian. Vector unparticle $A^\mu_\U$ can have non trivial $U(1)_Y$
quantum number, but cannot have non trivial $SU(3)_C$ and $SU(2)_L$ quantum numbers.

Let us study a simple example, involving two unparticle operators. Consider a charged scalar $S^+$
decaying into a charged scalar unparticle $\phi^+_\U$ of dimension $d_+$ and neutral scalar
unparticle $\phi^0_\U$ of dimension $d_0$. Under the SM gauge group $SU(3)_C\times SU(2)_L\times U(1)_Y$, $S^+$,
$\phi^+_\U$ and $\phi^0_\U$ transform as (1,1,2), (1,1,2) and (1,1,0), respectively. The lowest dimension
interaction possible is given by
$L_{eff} = (\lambda/\Lambda_\U^{d_+ + d_0 -3}) S^+ \phi^-_\U \phi^0_\U$.
We have decay distribution $d\Gamma$ for $S^+(p)\to \phi^+_\U(p_+)\phi^0_\U(p_0)$ given by
\begin{eqnarray}
d\Gamma(S^+\to \phi^+_\U \phi^0_\U) &=& \left (\lambda\over \Lambda_\U^{d_+ + d_0 -3}\right )^2{1\over 2 m_S}
(2\pi)^4 \delta^4(p - ( p_{+}+ p_{0}))\nonumber\\
&\times& A_{d_+} (p_{+}^2)^{d_{+} - 2}\theta(p^2_+) {d^4p_{+}\over (2\pi)^4}A_{d_0}
(p_{0}^2)^{d_{0} - 2}\theta(p^2_0) {d^4p_{0}\over (2\pi)^4}\;,
\end{eqnarray}
which leads to the energy distribution for the decay,
\begin{eqnarray}
{d\Gamma(S^+\to \phi^+_\U \phi^0_\U) \over d E_+}
&=& {|\lambda|^2\over \Lambda_\U^{2d_+ + 2d_0 -6}}{1\over 16\pi^3 m_S}A_{d_+}A_{d_0}E_+^{2d_+ -1}\nonumber\\
&\times& \int^{x_{max}}_0
x^{1/2}(1-x)^{d_+-2}(m_S^2 +E^2_+(1-2{m_S\over E_+} - x))^{d_0-2} dx\;.
\end{eqnarray}
Here $x = |\vec p_+|^2/E^2_+$. The limit for $E_+$ and $x$ are determined by energy momentum
conservation $p = p_+ + p_0$,
and also $p^2_+ >0$ and $p_0^2 >0$. We have: $x_{max} = 1$ for $0< E_+ < m_S/2$, and
$x_{max} = (1-m_S/E_+)^2$ for $m_S/2 < E_+ < m_S$.

Experimental signature would be a charged particle decay into a charge which can be
detected by measuring the energy deposited in the path plus missing energy.
The actual detectability depends on the scale $\Lambda_\U$ and the coupling $\lambda$.
Here our emphasis is on the different features compared with other processes.
If one only look at the charged track without energy measurement, there are several other
possibilities. For example: i) a usual charged particle decays into a lighter charged
particle plus a usual neutral undetected particle; or ii) a usual charged particle
decays into a neutral unparticle (particle) and a charged usual particle (unparticle).
If energy distributions of the charged track are measured one can distinguish different
scenarios. The possibility i) can be easily distinguished because the daughter charged
particle has a fixed energy. The possibility ii) can also be distinguished because the
charged track energy distribution is different from that for the two unparticle decay
discussed above. For example consider $S^+ \to \phi^+ \phi^0_\U$, here $\phi^+$ is a
usual charged scalar particle. The lowest dimension interaction is
$(\lambda/\Lambda_\U^{d_0 - 2}) S^+ \phi^- \phi^0_\U$
which leads to a differential energy distribution of the charged scalar particle given by
\begin{eqnarray}
{d\Gamma(S^+\to \phi^+ \phi^0_\U) \over d E_+} &=&
{|\lambda|^2\over \Lambda_\U^{2d_0 -4}}{1\over 8\pi^2 m_S}A_{d_0}
(m^2_S + E^2_+ - 2 m_S E_+ + m^2_+)^{d_0-2}(E^2_+ -
m^2_+)^{1/2}\;,
\end{eqnarray}
where $m_+$ is the mass of the charged scalar particle. The range
of $E_+$ is from $m_+$ to $(m_S^2-m^2_+)/2 m_S$. This differs from
the two unparticle case in Eq. (38), and is amenable to
experimental test.

In summary we have proposed a different approach to construct
unparticle operators based on scale invariant theories of
continuous mass. One can define unparticle like local operators
that couple to the SM. The unparticle properties emerge through
choice of interactions. There is no fixed point or dimensional
transmutation. The theory leads to clear understanding of how
unparticle exchange and phase space in the decay of SM particles
arises. We have generalized interactions of the standard model to
multiple unparticles in this formalism and have worked out some
examples for illustration. We show that products of unparticles are properly
normalized unparticles of dimension equal to the sum of the dimension of the
individual unparticles. We have extended our formalism to calculate three point functions
of unparticles. This required considering interactions of continuous mass fields.

\vspace*{0.2cm} {\bf Acknowledgement}: We thank T. Rizzo for
pointing out an error for the limit of x in eq. (21) in our
earlier manuscript. We thank Oyvind Tafjord for help with the nontrivial integral
in Eq. (23). XGH thanks the KITPC for hospitality while part of
this work was done. This work was supported in part by Grant No.
DE-FG02-96ER40949 of the Department of Energy, and the NSC and NCTS of
ROC.

\end{document}